# Fluid phases of argon

Leslie V. Woodcock

Department of Physics
Kyonggi University
Suwon
S. Korea

**A phase diagram of argon based upon percolation transition loci determined from literature experimental $\rho(p)$ isotherms, and simulation values using a Lennard-Jones model shows three fluid phases. The liquid phase spans all temperatures, from a metastable amorphous ground state at 0K, to ultra-high T. There is a supercritical mesophase bounded by two $2^{nd}$-order percolation transition loci, and a gas phase. Intersection of two percolation loci in the p-T plane thermodynamically defines an equilibrium line of critical density states.**

When van der Waals wrote his renowned Thesis [1] on the theory of the critical point in 1873, he would not have been aware of Gibbs work on thermodynamic equilibria, also published in 1873 [2]. The van der Waals critical point does not comply with Gibbs phase rule. Its existence is based upon a hypothesis rather than a thermodynamic definition. Moreover, no one has ever succeeded in measuring a critical density of an atomic or molecular fluid directly; liquid-gas critical densities are only obtained experimentally indirectly by an extrapolation of a mean of the two coexisting densities of liquid and vapor using the law of rectilinear diameters. Yet, the existence of a critical point singularity in the $\rho(p,T)$ density surface has not been questioned until very recently [3].

Gibbs defined surfaces of thermodynamic state functions, and subsequently explained all $1^{st}$-order thermodynamic phase transitions. A point on the $\rho(p,T)$ surface can only be defined thermodynamically where two lines intersect. For example, in a single phase region, a state point with two degrees-of-freedom (F) is the intersection of an isobar and an isotherm. A state point in a two-phase region (F=1), is the intersection of either an isotherm or an isobar with a coexistence line. There are no conceptual problems with the definition of the triple-point; it is the intersection of the liquid-vapor coexistence line and the solid-vapor coexistence line in the p-T plane. By contrast, there is no thermodynamic definition of the "critical point" of van der Waals; indeed if at $T_c$ $(dp/d\rho)_T = 0$, there is one phase and one degree of freedom; a contradiction of the laws of thermodynamics by Gibbs deductions.

This conundrum has been resolved from simulation studies on hard-sphere and square-well fluid percolation transitions [3,4]; at the critical temperature there is a 2-phase coexistence line between the densities of two percolation transitions. By analogy with the triple point, $T_c, p_c$ is found to be a "double point" with a single degree of freedom in the p-T plane where the percolation transition pressure loci intersect, as shown in **figure 1** for fluid argon. At $T_c$, each state-point corresponds to a different density, and since $(Vdp)_T = d\mu = 0$ there is a connecting line of states at $T_c, p_c$ of constant Gibbs chemical potential ($\mu$). Thus, the 2$^{nd}$-order percolation transitions on intersection in the p-T plane, become a 1$^{st}$-order phase transition, in compliance with Gibbs phase rule.

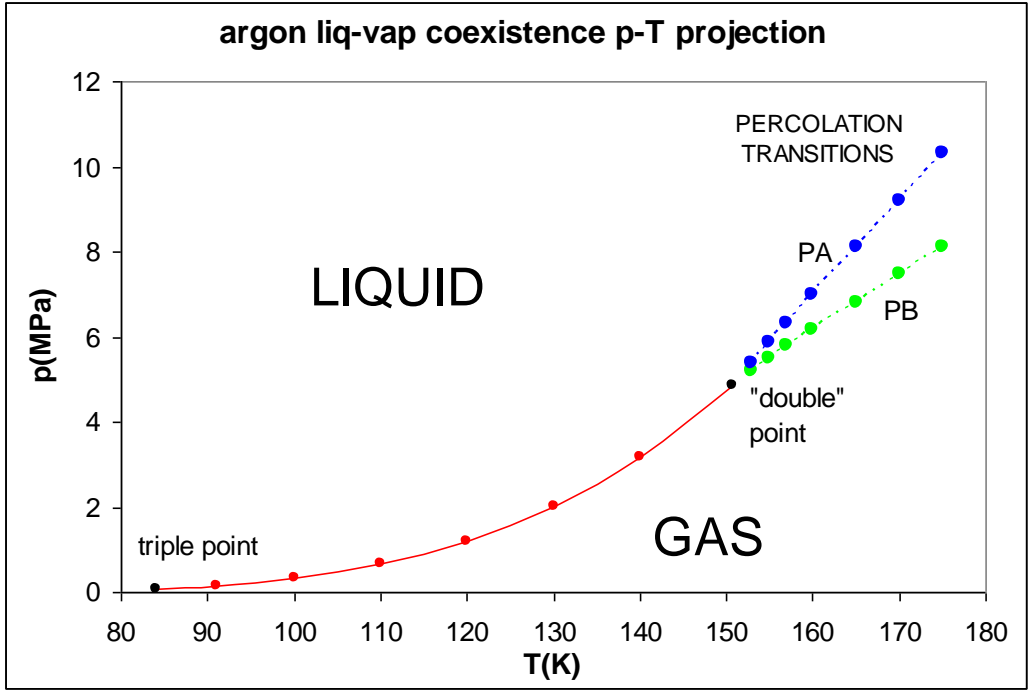

Figure 1 The liquid-vapor coexistence line for argon using the measurements of Gilgen et al. [6] for the coexistence line (red); PA (blue) is the available-volume percolation transition and PB (green) is the bonded-cluster percolation transition, both also obtained from the Gilgen supercritical experimental measurements [5]; the percolation transitions define the bounds of a supercritical mesophase.

The liquid-vapor coexistence line in **figure 1** is taken from the experimental p-V-T data of Gilgen et al. [5,6] which, except in the immediate vicinity of the critical temperature, is reported with 6-figure accuracy. The data points for the two percolation transitions are obtained from the discontinuity in the slope of the $p(\rho)$ isotherms, for seven supercritical isotherms as shown in **figure 2**. Using this information, and knowledge of the percolation transitions for the high-temperature limit of argon, if it is represented by a Lennard-Jones model, we can now obtain a reasonably accurate prediction the whole fluid phase diagram.

The available volume ($V_a$) percolation transition (PA) occurs at the density ($\rho_{pa}$) at which the volume accessible to any single mobile atom, in static equilibrium configuration of all the other atoms, percolates the whole system. It is related to the Gibbs chemical potential ($\mu$) by the equation

$$\mu = -k_B T \log_e (V_a / V) \qquad (1)$$

The bonded-cluster percolation transition (PB), is the same percolation transition, previously referred to for square-well fluids [3] as the extended-volume percolation transition. Now, for real molecules we need to distinguish between the excluded volume percolation transition (PE), which is also a cluster system-spanning transition, and the bonded-cluster percolation transition (PB).

At percolation transitions, thermodynamic state functions can change form due to sudden changes in state-dependence of density and/or energy fluctuations. For the hard-sphere fluid, PA is a very weak, but definite, higher-order thermodynamic phase transition [4]. Purely repulsive potential models have a gas-like region and a liquid-like region on either side of PA. The percolation transitions, as shown in **figure 1**, have been computed from simulations in hard-sphere and square-well model fluids, and have been found to play a central role in the thermodynamic description of liquid-gas critical phenomena. [3]

We do not presently know whether PE has a thermodynamic status for the hard-sphere fluid. When an attractive perturbation is added, however, both percolation transitions gain strength as temperature is reduced. This gives rise to $2^{nd}$-order thermodynamic phase transitions, in which there are discontinuities in second derivatives of chemical potential with respect to temperature or pressure, notably: isothermal compressibility $(d_2\mu/dp^2)_T$, heat capacity $(d_2\mu/dT^2)_p$ and thermal expansivity $(d_2\mu/dpdT)$ all of which undergo some degree of change at a percolation transitions.

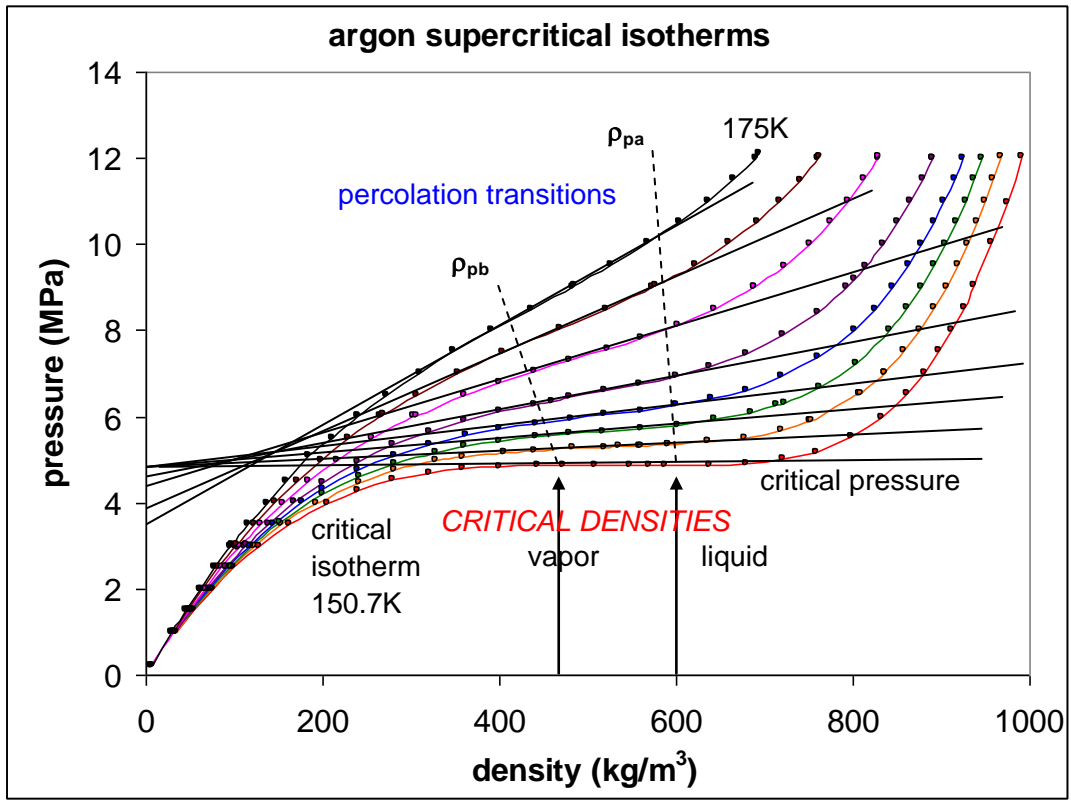

Figure 2: Experimental data points for the p-V isotherms of supercritical argon from the tabulations of Gilgen et al.[5]. The extended straight lines, fitted in the intermediate linear region, have been superimposed to highlight the percolation transition loci shown as dashed lines. $\rho_{pb}$ is the density of the bonded cluster percolation transition, and $\rho_{pa}$ is the density of the available-volume percolation transition.

The bonded-cluster percolation transition (PB) occurs when atoms bonded together, within a given characteristic distance, around the minimum in the pair potential, first begins to span the system. Unlike PA, PB manifests itself more in the temperature derivatives of the chemical potential which are determined by fluctuations in the energy, rather than density. Thus we are more likely to see lines of discontinuity, showing apparent maxima or minima, in the $2^{nd}$- order properties heat capacity and thermal expansivity.

The phase diagram in **figures 1 and 2** shows only the two percolation transitions that determine the critical coexistence. Atomic fluids have only one available-volume percolation transition (PA), but of the other two percolation transitions, the bonded-cluster percolation transition (PB) occurs at a much higher density than the excluded volume percolation transition (PE). Thus the bonded-cluster percolation intersects the available volume percolation line first at the higher temperature and pressure to effect the first-order phase transition as shown in **figure 1**. At present, we do not have information on the excluded volume percolation transition and its effect, if any, at lower temperatures, i.e. on intersecting the vapor-phase coexistence line at a temperature between $T_c$ and the triple point.

To obtain a general simple-fluid phase diagram, we use the Lennard-Jones potential with the scalable energy ($\varepsilon$) and distance corresponding to the diameter of a hard sphere reference fluid where $r_0$ is in dimension of the distance of zero force at $\varepsilon = -1$.

$$\phi(r) = \varepsilon\,[(r/r_0)^{-12} - 2(r/r_0)^{-6})] \qquad (2)$$

The phase diagram in Fig. 3 has been constructed follows. Beginning with the raw data from the Gilgen tables [5,6], the experimental mass densities at $T_c$ can be converted to reduced number densities for comparison with the known percolation transitions of the hard-sphere reference fluid. Using a Lennard–Jones pair potential argon take $\sigma = 3.405 \times 10^{-10}$m; then the reference hard-sphere diameter corresponding to zero-force is $r_0 = 2^{(1/6)}\,\sigma = 3.822 \times 10^{10}$m; taking Avagadro's number N= $6.0228 \times 10^{23}$ and argon atomic mass = 39.948, the data points of Gilgen have been converted to L-J units, with reduced number density $\rho = Nr_0^3/V$.

Nobody has ever measured a critical density directly; this is well illustrated from the experimental measurements of the argon liquid vapor coexistence densities by Gilgen et al. [6]. The highest temperature for which they report both coexisting vapor and liquid densities is 150.61. They use the law of rectilinear diameters to obtain their "critical point" temperature 150.69, and critical density 535.6 kg/m$^3$. The mean of the two highest recorded liquid and vapor densities is 536. The lowest coexisting liquid mass density they report is 602 kg/m$^3$ giving $\rho_{pa}(T_c) = 0.507$. The highest vapor mass density they can observe near $T_c$ is 470 kg/m$^3$ which corresponds to $\rho_{pb} = 0.395$. The line of critical states connects these two points (**figures 2 and 3**)

We can calculate the characteristic bond-length that defines the bonded-cluster percolation transition from the hard-sphere model, i.e. in a first-order perturbation approximation, if the structure is not perturbed significantly by the attraction. From an EXCEL power-law trendline parameterization of the extended volume percolation transition density as a function of cluster-length $\lambda$ from table I in the paper of Heyes et al. [7], an inversion gives

$$\lambda = 1 + 0.0412\, \rho^{-1.617} \qquad (3)$$

After conversion to L-J units, substituting $\rho$ for the minimum liquid density of Gilgen et al. at the critical temperature, we obtain $\lambda_{pa}(T_c) = 1.185$, which is in the anticipated range of the Lennard-Jones attraction; slightly greater than the distance of maximum attractive force which, from the first derivative of equation (2), is $(13/7)^{1/6} = 1.1086$.

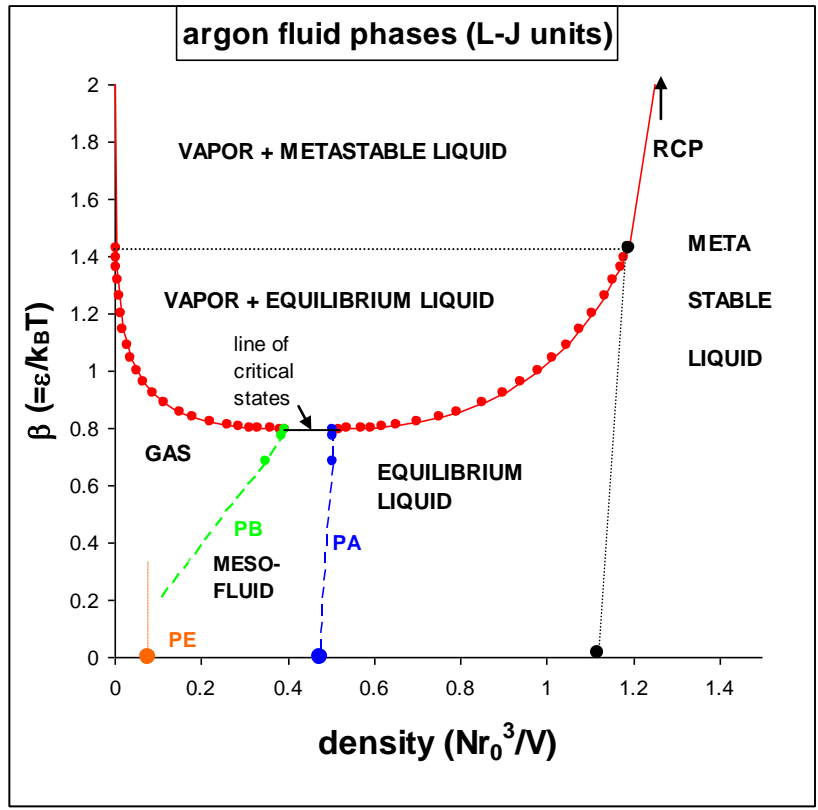

FIG 3: Phase diagram of argon, excluding crystalline phases, reduced to Lennard-Jones units; the red dots show the coexisting densities from the experimental measurements of Gilgen et al. [5,6]; the densities of the soft-sphere percolation transitions PE and PA in the high-temperature limit ($\beta=0$), and the soft-sphere freezing density (black dot) are obtained from references [7], [8] and [9] respectively.

To complete the construction of the phase diagram (**figure 3**) we need the two percolation densities $\rho_{pe}$ and $\rho_{pa}$ of the soft-sphere fluid which is the high-temperature limit of the L-J fluid. The density of PE for purely repulsive soft-spheres $\rho_{pe}$ (ss) = 0.085 is obtained from Heyes et al. [7]. The density of the soft-sphere available volume percolation density $\rho_{pa}$(ss) = 0.48 is determined in reference [8]. The equilibrium fluid freezing density of the high-temperature limit is obtained from Hoover et al. for the soft-sphere model [9], $\rho_f$(ss) = 1.51. These three phase transition points are the limiting high-temperature bounds of argon's three fluid phases at $\beta = \varepsilon/k_B T = 0$ in **figure 3**.

The bonded cluster percolation transition clearly must become increasingly weaker with temperature and eventually non-existent, probably before it crosses the excluded volume percolation transition at the much lower density. Likewise, it is not presently known what the extent and manifestations of the excluded volume percolation transition (PE) in figure 3 are at low temperatures. There is the possibility of a second mesophase bounded by PE and PB.

From the general picture in **Figure 3**, we can summarize the thermophysics of the liquid state. First, since a thermodynamic phase equation-of-state is continuous in all derivatives, within the F=2 Gibbs region of a phase diagram, the coexistence line will extend all the way from T=0 at an amorphous ground state ($\rho_a$). For argon at T→0K and vanishing pressures, this should be very close to the random close-packed density of the hard-sphere fluid $\rho_{rcp}$ = 1.216 [3]. The critical temperature ($T_c$) now has a thermodynamic definition; by the intersection of the available-volume and the bonded-cluster percolation transitions. In the p-T projection of the $\rho(p,T)$ Gibbs surface, these two lines intersect to trigger a first-order phase transition whereupon the gas at density $\rho_{pb}$ and liquid at density $\rho_{pa}$ coexist. The liquid phase extends all the way from 0K to an essentially-infinite temperature for a fluid defined by a classical model Hamiltonian, albeit metastable below the triple point temperature.

Brazhkin et al. [10] have calculated the values $C_p$, $\alpha_p$, and a "Widom line" (related to $\kappa_T$) for several isotherms of the Lennard-Jones supercritical fluid. These authors are unaware of the role of percolation transitions, but an inspection of their results is entirely consistent with the present thermodynamic based description. First, $C_p$ against pressure, in figure 3a of reference [10], shows a pronounced discontinuity at the reduced pressure $p\sigma^3/\varepsilon = 0.2$ (note: $\sigma = r_0/2^{(1/6)}$) for the lowest isotherm $T/T_c$ =1.4. Second $\alpha_p$ isotherms in figure 2b of reference [10] have flat maxima roughly spanning the two percolation densities shown here in **figures 2 and 3**. The isothermal compressibility ($\kappa_T$) increases monotonically from ideal gas to dense liquid, but when multiplied by a "correlation length", that decreases monotonically from ideal gas to dense liquid, the resulting maximum locus, which has been called the Widom line, seems to lie broadly within the two percolation densities $\rho_{pa}$ and $\rho_{pb}$ as seen in Figure 4 of reference [10].

Finally, we note that percolation transitions are known to be associated with discontinuities in linear and nonlinear transport properties [9], and various other dynamical properties such as frequency spectra [11]. There is an increasing literature [12] of hitherto "inexplicable" supercritical lines associated with observations of changes in dynamical properties from gas-like to liquid-like in various supercritical fluids, including water. It seems likely a simple thermodynamic explanation for the existence of all these lines will be forthcoming when the percolation transition loci in these fluids are investigated. Interestingly, the little TOC graphic presented within the abstract of the paper of Brazhkin et al. [10], shows all three lines of maxima stemming from the critical point in the p-T plane. All of their lines can be identified with the present thermodynamic percolation transition loci, shown here in figure 1 also in the p-T plane. Their $\zeta(T)$-max locus is near PB, their $C_p$-max locus is near PA, and their $\alpha_p$-max locus is intermediate between PA and PB loci. The line referred to by Brazhkin et al. [10] as a "Widom line" can actually be seen in Figure of the paper by Heyes and Melrose, published 25 years ago, and identified as the locus of a percolation transition [13].